\documentclass{sigchi-ext}
\usepackage[T1]{fontenc}
\usepackage{textcomp}
\usepackage[scaled=.92]{helvet} % for proper fonts
\usepackage{graphicx} % for EPS use the graphics package instead
\usepackage{balance}  % for useful for balancing the last columns
\usepackage{booktabs} % for pretty table rules
\usepackage{ccicons}  % for Creative Commons citation icons
\usepackage{ragged2e} % for tighter hyphenation

\def\plaintitle{Conversations On Multimodal Input Design With Older Adults} 
\def\emptyauthor{}
\def\plainkeywords{Human Computer Interaction; Older Adults; Interaction Design; Inputs}

\title{Conversations On Multimodal Input Design With Older Adults}

\numberofauthors{3}
\author{%
  \alignauthor{%
    \textbf{Adam S. Williams}\\
    \affaddr{Colorado State University} \\
    \affaddr{Fort Collins, CO 80525, USA} \\
    \email{AdamWil@colostate.edu} }
    \vfil \alignauthor{%
    \textbf{Sarah Coler}\\
    \affaddr{Columbine Health} \\
    \affaddr{Fort Collins, CO 80525, USA} \\
    \email{Sarah.Coler@columbinehealth.com}}
    \vfil \alignauthor{%
    \textbf{Francisco Ortega}\\
    \affaddr{Colorado State University} \\
    \affaddr{Fort Collins, CO 80525, USA} \\
    \email{fortega@colostate.edu } } }

\definecolor{linkColor}{RGB}{6,125,233}
\hypersetup{%
  pdftitle={\plaintitle},
  pdfauthor={\emptyauthor},
  pdfkeywords={\plainkeywords},
  bookmarksnumbered,
  pdfstartview={FitH},
  colorlinks,
  citecolor=black,
  filecolor=black,
  linkcolor=black,
  urlcolor=linkColor,
  breaklinks=true,
}

\begin{document}

%% For the camera ready, use the commands provided by the ACM in the Permission Release Form.
\CopyrightYear{2020}
\setcopyright{rightsretained}
\conferenceinfo{CHI'20,}{April  25--30, 2020, Honolulu, HI, USA}
\isbn{978-1-4503-6819-3/20/04}
\doi{https://doi.org/10.1145/3334480.XXXXXXX}
%% Then override the default copyright message with the \acmcopyright command.
\copyrightinfo{\acmcopyright}

\maketitle

% Uncomment to disable hyphenation (not recommended)

\RaggedRight{} 

% Do not change the page size or page settings.
\begin{abstract}

Multimodal input systems can help bridge the wide range of physical abilities found in older generations. After conducting a survey/interview session with a group of older adults at an assisted living community we believe that gesture and speech should be the main inputs for that system. Additionally, collaborative design of new systems was found to be useful for facilitating conversations around input design with this demographic. 

\end{abstract}


% ACM Classfication



% Print the classficiation codes

\section{Introduction}

% Creating intuitive input devices for older adults is of critical importance. The World Health Organization estimates that the population of elderly people (persons over 60) in the world will reach 2  billion by 2050. This represents nearly double the proportion of the population in that demographic since 2015 (12\% to 22\%)~\cite{WHO}.

Input technologies can help enable seniors to maintain their independence into later years by offloading some of the needs of daily activities, or by reducing the cognitive load needed for those activities~\cite{ROD86}. There are however differences in how older adults and younger adults interact with and perceive controls~\cite{CHU+12}. Some of these differences include perception limitations or inexperience with new technologies. 

%  Even though improper interface design was a contributing factor to interaction failure~\cite{CZA+19}.
Sadly, when interactions with a system are poor, older adults often assign personal blame instead of considering that the interface may not be properly designed for them~\cite{CHU+12}. When input technologies are not accessible it promotes feelings of exclusion and loss of control which can contribute negatively to one’s life~\cite{ROD86}.

% Gestures have shown promise as an appropriate interaction technique for older adults~\cite{KOB+11}.
Gestures have shown promise as an appropriate interaction technique for older adults, showing no significant difference in accuracy between older and younger adults~\cite{STO+09}. Speech has also been seen as a preferred input for older adults~\cite{BRU+01}.

\begin{margintable}[1pc]
  \begin{minipage}{\marginparwidth}
    \centering
    \begin{tabular}{r r}
      {\small \textbf{Scale}}
      & {\small \textbf{Response}} \\
      \toprule
      0 & Never \\
      1 & Once \\
    2 & Yearly \\
    3 & Monthly \\
    4 & Weekly \\
    5 & Daily \\
      \bottomrule
    \end{tabular}
    \caption{Response scale for device usage frequency}~\label{tab:Freq}
  \end{minipage}
\end{margintable}

\begin{margintable}[1pc]
  \begin{minipage}{\marginparwidth}
    \centering
    \begin{tabular}{r r}
      {\small \textbf{Scale}}
      & {\small \textbf{Response}} \\
      \toprule
      0 & Very Poor \\
      1 & Poor \\
    2 & Fair \\
    3 & Good \\
    4 & Excellent \\
      \bottomrule
    \end{tabular}
    \caption{Response scale for self assessment ratings}~\label{tab:self_ass}
  \end{minipage}
\end{margintable}
%Speech to has shown high potential as the input modality of choice for elderly and frail persons, with the potential to be more easily accepted than more intrusive solutions~\cite{POR+13}.

% These actions together constitute language~\cite{MCN05} and are fundamentally linked in both language production and comprehension~\cite{KEL+10}. 
%  Gesturing has been shown to help lower the cognitive load of a task~\cite{GOL+01} and lower task error rates~\cite{LEE+13}

Multimodal interfaces show promise in improving accessibility for older adults by providing a more natural and efficient interaction space~\cite{DAN+09}. Multimodal inputs are associated with higher user satisfaction, particularly with speech and gestural interfaces (touch)~\cite{JIA+13}. The benefits of this combined modality extend beyond user satisfaction. These input streams often contain non-redundant information which can lead to better recognizer systems development~\cite{KOO+93}.

\section{Survey}
To evaluate the feasibility of multimodal gesture and speech-based inputs we conducted a survey/interview session at an assisted living home. This pilot study is aimed at guiding our ideas about the future direction of input devices for this generation. This survey included questions about device usage frequency and comfort with using devices. The scales for responses are shown in tables \ref{tab:Freq} and \ref{tab:self_ass}. There were 11 participants (age M= 86.1, SD = 10.89). The range of ages spanned from 66 to 100 (8 male, 2 female). Previous careers ranged from homemaker to pilot.

% participants were given an informed consent, then individually a the questionnaires were administered. Questions were encouraged if any confusion was found. After all participants finished both forms a informal semi structured interview took place and lasted approximately 45 minutes. The survey was typed in 22 point font, no complaints about visibility were encountered. The participants ranged in physical ability, having various levels of hearing and vision loss. 

\section{Results}

Due to small sample sizes, we are treating the responses to the survey as a framing for the discussion. Thus, in-depth statistical analysis was not performed on the survey results. The results were impacted by age. People over 86 (group 2) selected ``never use'' for all of the device usage questions. People 86 and under (group 1) answered more of the usage questions. Most participants in group 1 report using a smartphone daily 3/4 (group 2 1/7). All phone usage questions were rated as daily or never. Only 2 participants reported using a mouse/keyboard daily, both in group 1.

\begin{marginfigure}[1pc]
  \begin{minipage}{\marginparwidth}
    \centering
    \includegraphics[width=0.9\marginparwidth]{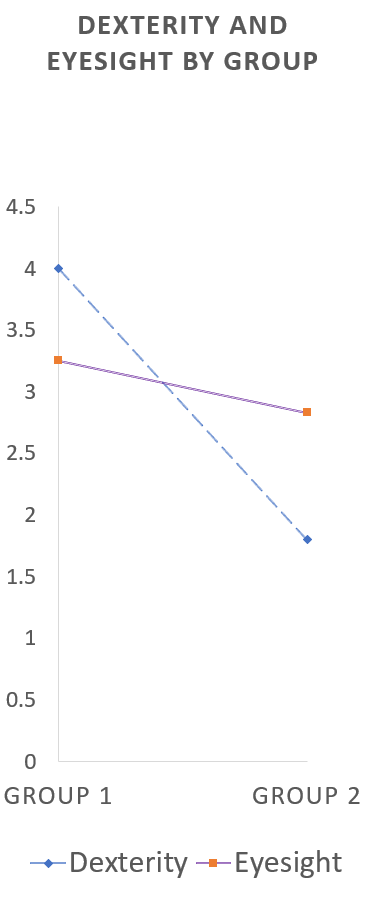}
    \caption{Participants self reported eyesight and manual dexterity by group}~\label{fig:Dex_Eye}
  \end{minipage}
\end{marginfigure}

Group 1 rated their eyesight as good (M=3.25), group 2 rated their eyesight as fair (M=2.83, SD=1.35). Dexterity followed the same trend. Group 1 rated their dexterity as good (M=4, SD=0), group 2 rated it as poor (M=1.88, SD=1.5). These results are shown in figure \ref{fig:Dex_Eye}. Group 1 rated their comfort with touchscreens as good (M=3, SD=.7). The mouse/keyboard and speech controls were rated as fairly comfortable both with (M=2, SD=1.41). For all device comfort questions group 2 rated their comfort as very poor.

%Daily remote use-age was listed for 2 people under 86 and one over. 2 people reported using a mouse and keyboard daily, both under 86. 

%One person reoprted using gusture controls monthly. One person used a Wii once and one person used a Virtual REality headset once. Both of these people were under 86. 

People said that they wanted to see more inputs that mirror what they currently use. Those being touchscreens. Some people felt that technologies are moving too fast. Indicating a preference for keeping inputs consistent in emerging technologies. When asked if they would like speech-based inputs people unanimously said no.  When asked if they thought using speech as commands for a system most people said it would be beneficial. The responses to these similar questions flipped based on the framing of the question.

After having difficulties getting input preferences for general new technologies we framed the questions around a virtual reality bingo game (VRBG). The participants helped design what this game would do. Having helped with design, participants were more involved in conversations around inputs for it. Bingo was chosen based on the advice of the activities director for the assisted living community. In their experience, the residents both enjoyed and could relate to the mechanics of bingo. This made the conversation relevant to the participants. Some participants gave more buy-in with the prospect of VRBG helping them to overcome physical limitations by scaling bingo cards to work around eyesight or dexterity limitations.

Participants wanted to touch the virtual bingo card to place their chips. A few people wanted speech-based inputs in combination with touch, but not as a stand-alone input. Participants said they wanted few controls in this system. One participant said ``simple is safe'' which was then repeated by other residents over the course of the conversation. 

% Outside of the scope of inputs, 
% They also wanted at minimum two bingo cards, with an option for more.

% \section{Discussion}

There was apprehension around participation in the conversation. Prior to the VRBG segment of the conversation, the thought of technology seemed to be off-putting to some participants. This was more pronounced in the participants above 86. Some participants, though eager to fill out the questionnaires and talkative during the initial session, become very quiet during the discussion about inputs. There was some level of guilt reported by several participants saying that they ``didn't want to ruin the results'' indicating that their self-perceived lack of experience with technology would be harmful to the survey. One participant went so far as to say they did not want to participate at all because they were ``computer illiterate'' and did not want to ruin the results. That participant later joined into the conversation. This looks similar to the self-blame found by Cui et al.~\cite{CHU+12}. Some other quotes from participants are found in table \ref{tab:Quotes}.

\section{Discussion}

The results from the survey and the interviews provide some direction for future input design. The interviews around input design also give insights on conducting meaningful conversations with this demographic. It was difficult to frame what we were asking when talking about what inputs they would like to see in future devices. There was a large disconnect between computers, phones, headsets, or any other type of technology. Different framings of the same question yielded drastically different answers.

% It was also necessary to give many examples of what different technologies are.

When considering hand-based inputs, gestures or otherwise, it's important to consider the level of mobility and hand usage older adults have. All of our participants reported having some level of manual dexterity decrease, with 4/11 indicating it was severe. This decrease in manual dexterity has been linked to decreased selection task performance~\cite{JIN+07}. The addition of speech-based inputs would make for a more flexible system, able to better cater to individual needs than touch or gesture alone. 

% Even after much talk about how there were no wrong answers and the goal was to asses what would be helpful for them, guilt was still mentioned. The Bingo game framing seemed to help with this.

%  Even when asking about technologies they are known to use, such as a Wii bowling game, there were difficulties in getting participant buy in. 

Going into a conversation with specific questions was difficult. There is a high potential for the participants to not contribute. Establishing a dialog around a common notion was helpful. The common notion in this instance was the VRBG. Having a member of the actives staff involved was impactful, the familiar face helped ease some of the anxieties in the room. That staff member was also able to help re-frame questions in a way that participants were more open to or more able to answer. 
% Using a large font for survey questions helped remove some of the visibility issues.

We recommend conducting interviews with this population across multiple periods. Establishing a rapport would help with participant involvement in the conversation. Most residents mentioned guilt about ``ruining results.'' We believe that having multiple sessions will help lessen this guilt which will help get at more meaningful conversations.

\section{Future Work}

This lab plans on developing the VRBG as a tool to get engagement in conversations about input device design from this population. We plan to have reoccurring meetings to help enable the participatory design of the system. While the end result may not be hugely impactful to this audience. The effect of having a stream of suggestions and iterative design based on the needs of this population will be.

\section{Conclusion}

Touchscreens and more intuitive Natural User Interfaces can enable older adults to join the digital world~\cite{PIN+12}. It is important that inputs are designed so that current older adults, and generations set to enter that space can remain in the virtual world. 

\begin{margintable}[1pc]
  \begin{minipage}{\marginparwidth}
    \centering
    \begin{tabular}{c}
        \textit{``you don't want me} \\
        \textit{to participate,}\\
        \textit{I'm computer illiterate''} \\
        \\
        \hline
        \\
        \textit{``simple is safe''} \\
        \\
        \hline
        \\
        \textit{``technology} \\
        \textit{moves too fast''}\\
      \bottomrule
    \end{tabular}
    \caption{Quotes from participants}~\label{tab:Quotes}
  \end{minipage}
\end{margintable}

Gesture and speech-based inputs are the best future direction for interaction design with older adults. Touch-based inputs were favored by most participants. We believe that gestures will begin to replace touchscreen interactions and can utilize many of the same features as touch. While there was some hesitation around speech inputs, we believe that as technology improves speech will become more accepted. When considering the next generation of older adults, the pervasive use of speech based home assistants might change this preference from touch. Multimodal systems that can utilize multiple input streams, in particular gestures and speech, can provide a robust interaction space. One capable of overcoming the individual level difficulties incurred while aging by providing alternative input options. With the wide range of abilities found in older generations, this flexibility is critical to widespread accessibility.

% Modality preferences change on the social context they are used in~\cite{WAS+06}

\balance{} 

\bibliographystyle{SIGCHI-Reference-Format}
\bibliography{sample}

\end{document}